\documentclass[prl,twocolumn,showpacs,aps]{revtex4-1}

\usepackage{amsmath}
\usepackage{amssymb}
\usepackage{txfonts}
\usepackage{graphicx}
\usepackage{epsfig}
\usepackage{enumerate}
\usepackage[utf8]{inputenc}
\usepackage{ifpdf}
\ifpdf
\usepackage{epstopdf}
\fi

\usepackage[unicode]{hyperref}
\hypersetup{
	pdftitle={YMsolutions},
	pdfauthor={Nian, Qian},
	pdfsubject={YMsolutions},
	colorlinks=true,
	linkcolor=blue,
	citecolor=blue,
	filecolor=black,
	urlcolor=blue
}

\def\url#1{}

\usepackage{bm}
\usepackage{color}

\newcommand{\be}{\begin{equation}}

\newcommand{\ee}{\end{equation}}

\newcommand{\bea}{\begin{eqnarray}}

\newcommand{\eea}{\end{eqnarray}}

\begin{document}

\title{A Topological Way of Finding Solutions to Yang-Mills Equation}
\author{Jun Nian$^{1}$}
\email{nian@umich.edu}
\author{Yachao Qian$^{2}$}
\email{yachao.qian@alumni.stonybrook.edu}
\affiliation{$^1$ Leinweber Center for Theoretical Physics, University of Michigan, Ann Arbor, MI 48109, USA}
\affiliation{$^2$ Department of Physics and Astronomy, Stony Brook University, Stony Brook, NY 11794-3800, USA}

\begin{abstract}

We propose a systematic way of finding solutions to classical Yang-Mills equation with nontrivial topology. This approach is based on one of Wightman axioms for quantum field theory, which is referred to as form invariance condition in this paper. For a given gauge group and a spacetime with certain isometries, thanks to this axiom that imposes strong constraints on the general Ansatz, a systematic way of solving Yang-Mills equation can be obtained in both flat and curved spacetimes. In order to demonstrate this method, we recover various known solutions as special cases as well as produce new solutions not previously reported in the literature.

\end{abstract}

\maketitle

{\flushleft\bf 1.~Introduction}

Up to now the most accurate law of elementary particle physics is given by quantum gauge theories \cite{YM, Faddeev, VeltmantHooft}. To fully understand a quantum gauge theory, one needs to study the vacuum structure of the theory, and the first step is to find exact solutions to classical Yang-Mills equation. In Euclidean flat spaces, many nontrivial solutions were found \cite{WuYang, instanton, multi-instanton}, which later lead to important progress in physics \cite{Shuryak, Nekrasov} and mathematics \cite{ADHM, MassGap}. Besides the solutions in flat spacetimes,  there are also nontrivial solutions found in curved spacetimes \cite{Verbin-1, Verbin-2, Nielsen, Lechtenfeld-1, Lechtenfeld-2, Lechtenfeld-3}.

Instead of searching for the solutions case by case, we would like to develop a systematic way of finding exact solutions with nontrivial topology. The key is to notice that all classical solutions respect one of the Wightman axioms~\cite{Wightman}, which will be referred to as form invariance condition. The new formalism can be summarized as a two-step approach:
\begin{enumerate}[(i)]
\item Obtain a general Ansatz explicitly satisfying the form invariance condition.
\item Insert this Ansatz into Yang-Mills equation.
\end{enumerate} 
After these two steps, the usually highly involved nonlinear partial differential equation is greatly simplified, and for the spherically symmetric case it even becomes an ordinary differential equation. Applying this new approach, we have recovered all the known spherically symmetric solutions in Euclidean three and four dimensions \cite{YM-1}. We have also discovered new solutions in Euclidean ten dimensions with the gauge group $SO(10)$. Since this approach relies only on the basic symmetries and Wightman axioms of quantum field theory, it is valid in more general cases other than just the flat spacetimes. For curved spacetimes, all the known solutions in the literature and some new solutions can be found using this approach.

In this paper, we demonstrate that this new approach can be applied to Yang-Mills theories in both flat and curved spacetimes, with a gauge group $SU(2)$ or other groups with higher rank. We considered the Euclidean signature for most cases, but a time-dependent solution will be discussed in the Lorentzian dS$_4$ space.


{\flushleft\bf 2.~Form Invariance Condition}

In attempting to define quantum field theory with solid mathematical foundations, Wightman et al. have established a set of axioms \cite{Wightman}. One of them for the vector field $j_\mu (x)$ is as follows:
\be\label{eq:axiom}
  \Lambda_\mu\,^\nu (A^{-1})\, j_\nu (A x + a) = U(a, A)\, j_\mu (x)\, U(a, A)^{-1}\, ,
\ee
where $\Lambda_\mu\,^\nu$ is a representation of the Lorentz group,  $A$ is a Lorentz transformation on the vector $x$, and $U(a, L)$ is a unitary or anti-unitary operator on the Hilbert space. The meaning of this condition is that a stable field configuration should be physically invariant under the pair of Lorentz transformations. In the conventional terminology, the condition \eqref{eq:axiom} can be interpreted as that viewing a gauge field in a specific frame is equivalent to choosing a gauge.

More specifically, for a field $V_\mu$ in the vector representation of the Lorentz group, the axiom \eqref{eq:axiom} becomes
\be
  (O^{-1})_\mu\,^\nu \, V_\nu (O\, x) = V_\mu (x)\, .
\ee
Of course, a generic vector field is not Lorentz invariant. However, we emphasize that the left-hand side of this condition consists of not just one Lorentz transformation, but a pair of Lorentz transformations, one acting on the coordinates $x$ and the other acting on the vector field index $\nu$, and these two Lorentz transformations are inverse to each other. Hence, after this pair of Lorentz transformations, a generic vector field should be physically invariant.

A gauge field is a vector field with additional gauge symmetry. Hence, the axiom \eqref{eq:axiom} for a gauge field $A_\mu$ reads:
\be\label{eq:FIlocal}
  (O^{-1})_\mu\,^\nu \, A_\nu (O\, x) = V^{-1} \, A_\mu (x) \, V + V^{-1} \partial_\mu V\, ,
\ee
where $O_\mu\,^\nu$ denotes a Lorentz transformation, and $V$ stands for a gauge transformation. Again, the left-hand side of \eqref{eq:FIlocal} consists of a pair of Lorentz transformations that are inverse to each other, hence, \eqref{eq:FIlocal} should lead to a physically equivalent configuration, i.e. a configuration that can be obtained through a gauge transformation. As one of the Wightman axioms, the condition \eqref{eq:FIlocal} should be true for any classical gauge fields, including both Abelian and non-Abelian ones. In fact, as we will see, all the known 3d and 4d classical solutions with non-trivial topology (e.g. the 4d Yang-Mills instanton) satisfy the condition \eqref{eq:FIlocal}.

The condition \eqref{eq:FIlocal} has been studied in Ref.~\cite{Gu}, where it is proven that for a Lorentz transformation with only rigid parameters the condition \eqref{eq:FIlocal} can be simplified to
\be\label{eq:FI}
  \left(O^{-1} \right)_\mu\,^\nu\, A_\nu (O\, x) = V^{-1}\, A_\mu (x)\, V\, ,
\ee
which is referred to as the form invariance condition throughout this paper.

  The form invariance condition \eqref{eq:FIlocal} or \eqref{eq:FI} is a necessary condition for the classical solutions to field equations. However, it turns out that these conditions are not automatically satisfied by an arbitrary field configuration, instead, they impose strong constraints on field configurations. Notice that a solution satisfying the field equation has to satisfy the condition \eqref{eq:FIlocal} or \eqref{eq:FI}, but the reverse is not true. The reason is simple: among all the configurations that are form invariant, a solution is only the one that minimizes the action. In case of the non-Abelian gauge theory, the field equation is Yang-Mills equation, which is our focus in this paper. The fact that a solution to Yang-Mills equation also satisfies the condition \eqref{eq:FIlocal} or \eqref{eq:FI} has been shown and discussed in great detail in \cite{YM-1}, which will also be demonstrated in this paper with many examples. Roughly speaking, for a given space we can construct an Ansatz which satisfies the condition \eqref{eq:FIlocal} or \eqref{eq:FI}, however, the Ansatz has to take a specific form in order to become a solution. Hence, the equation of motion is more restrictive than the conditions \eqref{eq:FIlocal} and \eqref{eq:FI}.

We emphasize that quantum fluctuations do not need to satisfy the form invariance condition \eqref{eq:FIlocal} or \eqref{eq:FI}. Instead, the form invariance condition is a necessary condition for the classical solutions. For those configurations that satisfy the condition \eqref{eq:FIlocal} or \eqref{eq:FI} but are not classical solutions, they form a new class of fluctuations, which distinguish from the standard quantum fluctuations in the literature. More properties of this new class of fluctuations are studied in \cite{YM-1}. In this paper we only discuss how to use the condition \eqref{eq:FIlocal} or \eqref{eq:FI} to systematically find solutions with nontrivial topology.

To find nontrivial solutions to Yang-Mills equation, our strategy is to first construct an Ansatz obeying the condition \eqref{eq:FIlocal} or \eqref{eq:FI} and other symmetry principles. The obtained Ansatz does not yet solve Yang-Mills equation, but will greatly simplify the original one. We will focus on the condition \eqref{eq:FI} in this paper.

{\flushleft\bf 3.~General Procedure}

Let us first list the general procedure of our approach to finding nontrivial solutions, and we then will demonstrate it with several examples. The procedure reads:
\begin{enumerate}
\item Construct the general Ansatz that is Lie algebra valued and invariant under the pair of Lorentz transformations.

\item Impose the form invariance condition \eqref{eq:FI}. This requirement will restrict the undetermined factors in the Ansatz constructed above.

\item Insert the Ansatz satisfying the form invariance condition \eqref{eq:FI} into Yang-Mills equation, which will greatly simplify the original equation. In the presence of some additional symmetries (e.g. spherical symmetry), Yang-Mills equation can even become a simple ordinary differential equation.

\item (Optional) Introduce a topological term in the action, which does not affect the theory at quantum level but is proportional to the winding number and makes the nontrivial topology manifest. It also provides boundary conditions for the solutions to Yang-Mills equation.

\item Solve the simplified Yang-Mills equation under certain boundary conditions, both of which are obtained before. Even if in some cases it is difficult to find analytical solutions, this well-defined differential equation system will make it easier to systematically analyze the nontrivial solutions to Yang-Mills equation.

\end{enumerate}

{\flushleft\bf 4.~Flat Space with Gauge Group $SU(2)$}

We first briefly review how the general formalism works for the spherically symmetric solutions in the flat 3d and 4d Euclidean spaces \cite{YM-1}.

For the 3d case with a gauge group $SU(2)$, the Ansatz reads
\be\label{eq:3DAnsatz}
 A_\mu = p(r) \left(U^{-1} \partial_\mu U\right)\, , 
\ee
where
\be\label{eq:Def_tau}
  r \equiv \sqrt{x_\mu x_\mu}\, ,
\ee
and the index $\mu \in \{1, 2, 3\}$ labels the spacetime coordinates. $U$ is an $SU(2)$ group element. In general, $U$ can be expressed as
\be\label{eq:3DAnsatz1a}
 U = \textrm{exp} \left[T_a \theta^a (x) \right] = \textrm{exp} \left[T_a\, \omega^a\,_\mu\, \hat{n}_\mu \theta(r) \right]\, ,
\ee
where the index $a \in \{1, 2, 3\}$ labels the generators of $SU(2)$ and $\hat{n}_\mu \equiv x_\mu / r$. The tensor $\omega^a\,_\mu$ is an $O(3)$ group element.

For the 4d Euclidean space with a gauge group $SU(2)$, the Ansatz is
\be\label{eq:4DAnsatz}
  A_\mu = p(r, x_4) \left(U^{-1} \partial_\mu U\right)\, ,\quad U = \textrm{exp} \left[T_a\,\omega^a\,_i\, \hat{n}_i \, \theta(r, x_4) \right]\, ,
\ee  
where $r$ is formally defined in the same way as the 3d case \eqref{eq:Def_tau} but with the index $\mu$ running from $1$ to $4$, and the index $a \in \{1, 2, 3\}$ still labels the generators of $SU(2)$, while the index $i \in \{1, 2, 3\}$ stands for three out of four directions in 4d Euclidean space.  For the gauge group $SU(2)$, we choose the anti-hermitian generators as
\be\label{eq:generator}
  T^a = \frac{\sigma^a}{2 \textrm{i}}
\ee
with the standard Pauli matrices $\sigma^a$ and the imaginary unit i.

To simplify the Ansatz, we impose the form invariance condition \eqref{eq:FI}. For the 3d case, by considering the spatial rotations in $(12)$, $(23)$ and $(31)$ planes, one can show that the tensor $\omega^a\,_\mu$ is constrained to be a constant $O(3)$ group element~\cite{YM-1}. For simplicity, we choose it to be $\delta^a\,_\mu$, then the Ansatz \eqref{eq:3DAnsatz} automatically satisfies the condition~\eqref{eq:FI}, which implies that for the 3d case there is no constraint on the factors $p(r)$ or $\theta(r)$. For the 4d case, however, due to the extra spatial rotations in $(14)$, $(24)$ and $(34)$ planes, it was proven in Appendix~E of Ref.~\cite{YM-1} that the form invariance condition \eqref{eq:FI} requires additionally $p(r, x_4) = p (r)$ and fixes the function $\theta(r, x_4)$ to be
\be\label{eq:4Dq}
  \textrm{cos} \frac{\theta}{2} = \pm \frac{x_4}{r}\, ,
\ee
where the signs are related by a gauge transformation, and we pick up the positive sign. Consequently, the 4d Ansatz \eqref{eq:4DAnsatz} can be expressed as
\be\label{eq:amuansatz4D}
  A_\mu^a\, T^a \,  =  \, 2 \frac{p(r)}{r^2}\eta_{a\mu\nu} x_\nu T^a\, ,
\ee
where $\eta_{a\mu\nu}$ are the 't Hooft symbols.

As we have shown here, the form invariance condition strongly restricts the Ansatz. Let us emphasize this point by another example. For instance, one may write an expression with two types of 't Hooft symbols:
\be\label{eq:AnsatzExample}
  C_1 \eta_{a \mu \nu} x_\nu T^a + C_2 \bar{\eta}_{a \mu \nu} x_\nu T^a\qquad (C_{1, 2}: \textrm{ constants})
\ee
as an Ansatz to 4d Yang-Mills equation, but in general it is only form invariant for the gauge group $SO(4)$, but not for $SU(2)$. In order for the expression \eqref{eq:AnsatzExample} with the gauge group $SU(2)$ to be form invariant, either $C_1$ or $C_2$ has to vanish. Hence, the form invariance condition \eqref{eq:FI} imposes very strong constraints on the Ansatz in general.

{\flushleft\bf 5.~Topological Term and Boundary Condition}

Inserting the 3d Ansatz \eqref{eq:3DAnsatz} back into Yang-Mills equation, we obtain the greatly simplified equation, which in this case is just an ordinary differential equation. Before solving the simplified Yang-Mills equation, it will be more convenient if we introduce a topological term to fix the boundary conditions. The 3d topological term is the Chern-Simons term:
\be\label{eq:S_CS}
   S_{CS} = \frac{\textrm{i}k}{4\pi} \int \textrm{d}^3 x\, \textrm{Tr} \left(A \wedge \textrm{d}A + \frac{2}{3} A\wedge A\wedge A \right) \equiv 2 \pi \textrm{i} k B\, ,
\ee
where $k$ is the Chern-Simons level, and $B$ denotes the winding number, both of which should be integers.

Substituting the 3d Ansatz \eqref{eq:3DAnsatz} into \eqref{eq:S_CS}, it is easy to show that the winding number $B$ has contributions only from the singular points of the integrand, which reads
\be 
  B = \frac{3}{2\pi^2} \sum_\beta \left(\frac{2}{3} p_\beta^3 - p_\beta^2 \right) \left(\theta_\beta - \frac{1}{2} \textrm{sin} 2 \theta_\beta \right) \int \textrm{d}S_\beta \, \hat{n}\cdot (\partial_1 \hat{n} \times \partial_2 \hat{n})\, ,
\ee
where $\beta$ denotes the singular points, e.g., $r = 0$ and $r =\infty$ for the simplest cases, which satisfy
\be
\frac{1}{4\pi} \int \textrm{d}S  \, \hat{n}\cdot (\partial_1 \hat{n} \times \partial_2 \hat{n}) = \pm 1 \, ,
\ee
where the contributions from $r=0$ and $r = \infty$ have the opposite sign due to different boundary orientations.

 By requiring that the Chern-Simons term provides us with a well-defined winding number, we see that the values of $p(r)$ and $\theta(r)$ at the boundaries $r = 0$ and $\infty$ are fixed to be
\be\label{eq:3Dbc}
  p \big|_{r = 0,\, \infty} = \frac{1}{2}\, \mathbb{Z}_{\geq 0}\, ,\quad \theta \big|_{r = 0,\, \infty} = \pi\, .
\ee
For the 4d case, we can introduce
\be\label{eq:S_top}
  S = - \frac{\textrm{i}}{8 \pi} \int \textrm{d}^4 x\, \textrm{Tr} \left[F_{\mu\nu} \left(* F_{\mu\nu} \right) \right] \equiv 2 \pi \textrm{i} B\, ,
\ee
where $B$ denotes the 4d winding number.

 Applying the 4d Ansatz \eqref{eq:4DAnsatz} to \eqref{eq:S_top}, we obtain
\be
  B = - \left(2 p^3  - 3 p^2  \right) \mathcal{T}\big{|}_{r = 0}   - \left(2 p^3  - 3 p^2  \right) \mathcal{T}\big{|}_{r = \infty}\, ,
\ee
where
\begin{align}
  {} & \mathcal{T} \big{|}_{r = 0} \nonumber\\
  = & \,  \frac{1}{24 \pi^2} \int_{r = 0} \textrm{d}\Omega_\mu \, \epsilon_{\mu\nu\rho\sigma} \, \textrm{Tr} \left[ \left( U^{-1} \partial_\nu U \right) \left( U^{-1} \partial_\rho U \right) \left( U^{-1} \partial_\sigma U \right) \right] \nonumber\\
  = & \, 1
\end{align}
for the gauge group $SU(2)$. Due to the opposite boundary orientations, the contributions from $r=0$ and from $r = \infty$ have opposite signs.

By requiring a well-defined winding number, we find that the values of $p(r)$ at the boundaries $r = 0$ and $\infty$ have to be
\be\label{eq:4Dbc}
  p \big|_{r = 0,\, \infty} = \frac{1}{2}\, \mathbb{Z}_{\geq 0}\, .
\ee

Finally, we obtain the simplified Yang-Mills equation in 3d and 4d with the boundary conditions \eqref{eq:3Dbc} and \eqref{eq:4Dbc} respectively. The solutions with the lowest winding numbers are listed as follows. For the 3d case, we find
\begin{itemize}
\item Vacuum solution:
\be\label{solution3DVac}
  A_{\mu, a} = 0\, .
\ee

\item Wu-Yang monopole:
\be\label{solution3DWYM}
  A_{\mu, a} = - \frac{\epsilon_{\mu a i} x_i}{r^2}\, .
\ee

\item Pure gauge solution:
\be\label{solution3DPGS}
  A_{\mu, a} = - 2 \frac{\epsilon_{\mu a i} x_i}{r^2}\, .
\ee
\end{itemize}
For the 4d case, we can recover all the known solutions in the literature as follows:
\begin{itemize}
\item Vacuum solution:
\be
  A_{\mu, a} = 0\, .
\ee

\item Meron solution:
\be
  A_{\mu, a} = \frac{1}{r^2}\, \eta_{a\mu\nu} x_\nu\, .
\ee

\item Instanton solution:
\be
  A_{\mu, a} = \frac{2}{r^2 + c}\, \eta_{a\mu\nu} x_\nu\, ,
\ee
with a positive real constant $c$.

\item Anti-instanton solution:
\be
  A_{\mu, a} = \frac{2 c}{r^2 (r^2 + c)}\, \eta_{a\mu\nu} x_\nu\, ,
\ee
with a positive real constant $c$.

\item Pure gauge solution:
\be
  A_{\mu, a} = \frac{2}{r^2}\, \eta_{a\mu\nu} x_\nu\, .
\ee
\end{itemize}

{\flushleft\bf 6.~3d Curved Space with Gauge Group $SU(2)$}

Since Yang-Mills equation is classically conformal invariant, the solutions in conformally-flat curved spacetimes can be easily derived by our approach in the same way as their corresponding flat-space solutions. Here, we would like to illustrate that our approach can be used in more general cases.

Let us consider the classical solutions to the Yang-Mills equation on the 3d curved space with spherical symmetry for simplicity. Because it shares the same isometry with the 3d flat space, they can be treated in a similar way. More explicitly, the 3d curved space is given by the metric:
\be\label{eq:3dMetricSphericalSymm}
  \textrm{d}s^2 = \textrm{d}r^2 + f(r)\, \textrm{d}\Omega_{S^2}^2\, ,
\ee
where $\textrm{d}\Omega^2_{S^2}$ denotes the line element on the 2-sphere $S^2$, which can be embedded into a 3d flat Euclidean space, and $r$ is the radial coordinate, which is related to the 3d flat Euclidean space coordinates in the standard way:
\be
  r^2 \equiv |x|^2 = x_1^2 + x_2^2 + x_3^2\, .
\ee
Hence, for this class of curved spaces there is an $SO(3)$ rotational isometry acting on the vector $(x_1,\, x_2,\, x_3)^T$.

Now consider the Ansatz for the Yang-Mills field:
\be\label{eq:3DCurvedAnsatz}
  A_\mu = p(r)\, U^{-1} \partial_\mu U\quad \textrm{with}\quad U = \textrm{exp} \Bigg[T_a\, \omega^a\,_i\, \frac{x^i}{|x|} \theta(r) \Bigg]\, .
\ee
It is a generalization of the Ansatz \eqref{eq:3DAnsatz}, and the only difference is that \eqref{eq:3DCurvedAnsatz} is written in terms of the flat space coordinates, in which the $S^2$ is embedded. Notice that \eqref{eq:3DCurvedAnsatz} is a legitimate Ansatz for these curved spaces, because the spaces with the metric \eqref{eq:3dMetricSphericalSymm} have the same isometry $SO(3)$ as the flat space $\mathbb{R}^3$. Hence, all the previous analyses for the flat $\mathbb{R}^3$ still hold for the class of the curved spaces with the metric \eqref{eq:3dMetricSphericalSymm}. In fact, the flat space $\mathbb{R}^3$ is just a special case of the metric \eqref{eq:3dMetricSphericalSymm} with $f(r) = r^2$.

The metric \eqref{eq:3dMetricSphericalSymm} can be written as
\be
  \textrm{d}s^2 = \frac{f(r)}{r^2} \Bigg[\frac{r^2}{f(r)} \textrm{d}r^2 + r^2\, \textrm{d}\Omega^2_{S^2} \Bigg] = \frac{f(r)}{r^2} \Big[h(r) \textrm{d}r^2 + r^2\, \textrm{d}\Omega^2_{S^2} \Big]\, ,
\ee
i.e., it is conformally equivalent to
\be\label{eq:3dMetricSphericalSymm-2}
  \textrm{d}s^2 = h(r) \textrm{d}r^2 + r^2\, \textrm{d}\Omega^2_{S^2}\, .
\ee
We will focus on the metric \eqref{eq:3dMetricSphericalSymm-2} in the following, which can be written as
\be
  g_{\mu \nu} = \delta_{\mu \nu} + \frac{h(r) - 1}{r^2} x_\mu x_\nu\, ,
\ee
where we use pseudo-Cartesian coordinates --- the indices of the coordinates $x^\mu$ or $x_\mu$ are raised and lowered by the Kroneckar delta, not by the metric itself. Notice that $r^2 \equiv |x|^2$ preserves an $SO(3)$ rotational symmetry only when we treat $x^\mu$ and $x_\mu$ on the equal footing. The inverse metric is
\be
  g^{\mu \nu} = \delta^{\mu \nu} + \frac{1 - h(r)}{r^2\, h(r)} x^\mu x^\nu\, ,
\ee
and the Christoffel symbol is given by
\be\label{eq:Christoffel}
  \Gamma^\mu\,_{\nu \rho} = \frac{h(r) - 1}{r^2\, h(r)} g_{\nu \rho} x^\mu + \frac{r^2 h'(r) - 2 r (h(r) - 1)}{2 r^5 h(r)} x^\mu x_\nu x_\rho\, .
\ee

Because we are using the pseudo-Cartesian coordinates, the Ansatz $A_{\mu, a}$ satisfying the form invariance condition on the curved space \eqref{eq:3dMetricSphericalSymm-2} is formally the same as the one in the flat $\mathbb{R}^3$:
\be
  A_{\mu, a} = G \Bigg(\frac{\delta_{\mu a}}{r} - \frac{x_\mu x_a}{r^3} \Bigg) + (H - 1) \frac{\epsilon_{\mu a i} x^i}{r^2}\, .
\ee
Consequently,
\begin{align}
  F_{\mu \nu, a} & = \Bigg(\frac{x_\mu \delta_{\nu a} - x_\nu \delta_{\mu a}}{r^3} \Bigg) r G' + \frac{\epsilon_{\mu \nu a}}{r^2} (G^2 + 2 H - 2) \nonumber\\
  {} & \quad + \frac{x^i (x_\mu \epsilon_{\nu a i} - x_\nu \epsilon_{\mu a i})}{r^4} (2 - 2 H + r H' - G^2)\nonumber\\
  {} & \quad + \frac{x_a x^i \epsilon_{\mu \nu i}}{r^4} (H - 1)^2\, .
\end{align}
When we raise the indices, the expression is given by
\begin{align}
  F^{\mu \nu, a} & = \Bigg(\frac{x^\mu \delta^{\nu a} - x^\nu \delta^{\mu a}}{r^3} \Bigg) \frac{r}{h} G' + \frac{\epsilon^{\mu \nu a}}{r^2} (G^2 + 2 H - 2) \nonumber\\
  {} & \quad + \frac{x_i (x^\mu \epsilon^{\nu a i} - x^\nu \epsilon^{\mu a i})}{r^4} \left(2 - 2 H + \frac{r}{h} H' - G^2 \right) \nonumber\\
  {} & \quad + \frac{x^a x_i \epsilon^{\mu \nu i}}{r^4} (H - 1)^2\, ,
\end{align}
where
\be
  \frac{\partial}{\partial x_r} = g^{rr} \frac{\partial}{\partial x^r} = \frac{1}{g_{rr}} \frac{\partial}{\partial x^r} = \frac{1}{h(r)} \frac{\partial}{\partial r}\, 
\ee
with $x^r \equiv r$.

Finally, we obtain $D_\mu F^{\mu \nu, a}$ on the curved space \eqref{eq:3dMetricSphericalSymm-2}:
\begin{align}
  D_\mu F^{\mu \nu, a} & = \partial_\mu F^{\mu \nu, a} + \Gamma^\mu\,_{\mu \rho} F^{\rho \nu, a} + \Gamma^\nu\,_{\mu \rho} F^{\mu \rho, a} + \epsilon^{abc} A_{\mu, b} F^{\mu\nu}\,_c \nonumber\\
  {} & = \frac{x^a x^\nu}{r^5 h^2} \Bigg[ - G h^2 + G^3 h^2 + G h^2 H^2 + 2 r h H G' + \frac{1}{2} r^2 G' h' \nonumber\\
  {} & \qquad\qquad - 2 r h G H' - r^2 h G'' \Bigg] \nonumber\\
  {} & \quad + \frac{\epsilon^{\nu a i} x_i}{r^4 h^2} \Bigg[ h^2 H - G^2 h^2 H - h^2 H^3 - \frac{1}{2} r^2 h' H' + r^2 h H'' \Bigg] \nonumber\\
  {} & \quad + \frac{\delta^{a \nu}}{r^3 h^2} \Bigg[G h^2 - G^3 h^2 - G h^2 H^2 - \frac{1}{2} r^2 G' h' + r^2 h G'' \Bigg]\, .
\end{align}
Hence, the Yang-Mills equation $D_\mu F^{\mu \nu, a} = 0$ on the curved space \eqref{eq:3dMetricSphericalSymm-2} can be reduced to
\begin{align}
  - G h^2 + G^3 h^2 + G h^2 H^2 + 2 r h H G' + \frac{1}{2} r^2 G' h' & {} \nonumber\\
  - 2 r h G H' - r^2 h G'' & = 0\, ,\label{eq:3dYM-1}\\
  h^2 H - G^2 h^2 H - h^2 H^3 - \frac{1}{2} r^2 h' H' + r^2 h H'' & = 0\, ,\label{eq:3dYM-2}\\
  G h^2 - G^3 h^2 - G h^2 H^2 - \frac{1}{2} r^2 G' h' + r^2 h G'' & = 0\, .\label{eq:3dYM-3}
\end{align}
Combining Eqs.~\eqref{eq:3dYM-1} and \eqref{eq:3dYM-3}, we obtain a simple relation
\be
  H G' = G H'\, .
\ee
We distinguish four possibilities:
\begin{enumerate}
  \item $G = c H$, $G \neq 0$ and $H \neq 0$ ;
  \item $G \neq 0$, $H = 0$ ;
  \item $G = 0$, $H \neq 0$ ;
  \item $G = 0$, $H = 0$.
\end{enumerate}
For the first case, all three equations \eqref{eq:3dYM-1} - \eqref{eq:3dYM-3} lead to the same equation:
\be\label{eq:3dYM-4}
  H - (1 + c^2) H^3 - \frac{r^2 h'}{2 h^2} H' + \frac{r^2}{h} H'' = 0\, .
\ee

Firstly, we notice that the solutions in flat $\mathbb{R}^3$ remain as valid solutions:
\begin{itemize}
\item Vacuum solution:
\be\label{solution3DVac}
  G=0,\, H=1\quad\Rightarrow\quad A_{\mu, a} = 0\, .
\ee

\item Pure gauge solution:
\be\label{solution3DPGS}
  G=0,\, H=-1\quad\Rightarrow\quad A_{\mu, a} = - 2 \frac{\epsilon_{\mu a i} x_i}{r^2}\, .
\ee

\item Nontrivial solution:
\be\label{solution3DWYM}
  G=0,\, H=0\quad\Rightarrow\quad A_{\mu, a} = - \frac{\epsilon_{\mu a i} x_i}{r^2}\, .
\ee
\end{itemize}
Although it is not surprising that there are vacuum solution and pure gauge solution for any $h(r)$, it is interesting that \eqref{solution3DWYM} is always a solution, which corresponds to the Wu-Yang monopole in flat 3d space. As we have shown, it does not just solve  Yang-Mills equations on a specific curve space but on all curved spaces in the class \eqref{eq:3dMetricSphericalSymm-2}. To our best knowledge, these are new solutions to the 3d Yang-Mills equation on curved spaces and have not been discussed in the literature before.

Secondly, once we know the factor $h(r)$ in the 3d metric \eqref{eq:3dMetricSphericalSymm-2}, we can solve the simplified differential equations obtained from \eqref{eq:3dYM-1} - \eqref{eq:3dYM-3} to find more solutions to $G$ and $H$ and consequently more nontrivial solutions to the original Yang-Mills equation on the 3d curved spacetimes given by the metric \eqref{eq:3dMetricSphericalSymm-2}.

{\flushleft\bf 7.~4d Curved Space with Gauge Group $SU(2)$} 

Similar to the 3d case, let us consider the classical solutions to the Yang-Mills equation on the 4d curved space with $SO(4)$ rotational symmetry for simplicity. Because it shares the same isometry with the 4d flat space, it can be treated in a similar way.
 The 4d curved space is given by the metric: 
\be\label{eq:4dMetricSphericalSymm-1}
  \textrm{d}s^2 = \textrm{d}r^2 + f(r)\, \textrm{d}\Omega^2_{S^3}\, ,
\ee
where $\textrm{d}\Omega^2_{S^3}$ denotes the line element on the 3-sphere $S^3$, which can be embedded into a 4d flat Euclidean space. $r$ is the radial coordinate, which is related to the 4d flat Euclidean space coordinates in the standard way:
\be
  r^2 \equiv |x|^2 = \sum_{i=1}^4 x_i^2\, .
\ee
Hence, for this class of curved spaces there is an $SO(4)$ rotational isometry acting on the vector $(x_1,\, x_2,\, x_3,\, x_4)^T$.

In the following, instead of the 4d metric \eqref{eq:4dMetricSphericalSymm-1} we consider a conformally equivalent metric:
\be\label{eq:4dMetricSphericalSymm-2}
  \textrm{d}s^2 = h(r)\, \textrm{d}r^2 + r^2 \textrm{d}\Omega^2_{S^3} = \textrm{d}\vec{x}^2 + \frac{h (r) - 1}{r^2} (\vec{x} \cdot \textrm{d}\vec{x})^2\, ,
\ee
i.e., in the pseudo-Cartesian coordinates
\be
  g_{\mu \nu} = \delta_{\mu \nu} + \frac{h(r) - 1}{r^2} x_\mu x_\nu\, .
\ee

Now for the 4d curved spaces with the metric \eqref{eq:4dMetricSphericalSymm-2}, the Ansatz satisfying the form invariance condition \eqref{eq:FI} formally remains the same as \eqref{eq:4DAnsatz} in the flat $\mathbb{R}^4$ space. After taking into account the result \eqref{eq:4Dq} in $\mathbb{R}^4$ derived from the form invariance condition \eqref{eq:FI}, we obtain:
\be\label{eq:AnsatzCurved}
  A_{\mu,\, a} = 2 \frac{p (r)}{r^2} \eta_{a \mu \nu} x^\nu\, ,
\ee
where $\eta_{a \mu \nu}$ again denote the 't Hooft symbols. Consequently,
\begin{align}
  F_{\mu\nu, a} & = 4 \eta_{a \mu \nu} \Big(\frac{p^2}{r^2} - \frac{p}{r^2} \Big) + 4 x^\rho (x_\mu \eta_{a \nu \rho} - x_\nu \eta_{a \mu \rho}) \Big(\frac{p^2}{r^4} - \frac{p}{r^4} + \frac{p'}{2 r^3} \Big) \, ,\nonumber\\
  F^{\mu\nu, a} & = 4 \eta^{a \mu \nu} \Big(\frac{p^2}{r^2} - \frac{p}{r^2} \Big) + 4 x_\rho (x^\mu \eta^{a \nu \rho} - x^\nu \eta^{a \mu \rho}) \Big(\frac{p^2}{r^4} - \frac{p}{r^4} + \frac{p'}{2 h r^3} \Big) \, .
\end{align}
Moreover, the Christoffel symbols for the metric \eqref{eq:4dMetricSphericalSymm-2} formally take the same expression as \eqref{eq:Christoffel}. Therefore,
\begin{align}
  D_\mu F^{\mu \nu,\, a} & = \partial_\mu F^{\mu \nu, a} + \Gamma^\mu\,_{\mu \rho} F^{\rho \nu, a} + \Gamma^\nu\,_{\mu \rho} F^{\mu \rho, a} + \epsilon^{abc} A_{\mu, b} F^{\mu\nu}\,_c  \nonumber\\
  {} & = \frac{8 \eta^{a \nu \rho} x_\rho}{r^4} \Big(- p + 3 p^2 - 2 p^3 - \frac{r^2}{8 h^2} p' h' + \frac{r}{4 h} p' + \frac{r^2}{4 h} p'' \Big)\, .
\end{align}
The Yang-Mills equation on the curved spaces \eqref{eq:4dMetricSphericalSymm-2} reduces to
\be\label{eq:4dYM}
  - p + 3 p^2 - 2 p^3 - \frac{r^2}{8 h^2} p' h' + \frac{r}{4 h} p' + \frac{r^2}{4 h} p'' = 0\, .
\ee

First of all,  we can easily see that for any $h(r)$ there are always the following solutions:
\begin{itemize}

\item Vacuum solution:
\be
  p = 0 \quad\Rightarrow\quad A_{\mu, a} = 0\, .
\ee

\item Pure gauge solution:
\be
  p = 1 \quad\Rightarrow\quad A_{\mu, a} = \frac{2}{r^2} \eta_{a \mu \nu} x^\nu\, .
\ee

\item Nontrival solution:
\be
  p = \frac{1}{2} \quad\Rightarrow\quad A_{\mu, a} = \frac{1}{r^2} \eta_{a \mu \nu} x^\nu\, .
\ee

\end{itemize}
Similar to the 3d case, besides the vacuum solution and pure gauge solution, there is always a nontrivial solution with $p = 1/2$. For the flat 4d space, it corresponds to the meron solution, but as we have shown here, such solution exists not just on a specific space but on all 4d curved spaces in the class \eqref{eq:4dMetricSphericalSymm-2} . To our best knowledge, these are new solutions to the 4d Yang-Mills equation on curved spaces and have not been discussed in the literature before.

Once we know the factor $h(r)$ in the metric \eqref{eq:4dMetricSphericalSymm-2}, by solving the differential equation \eqref{eq:4dYM} we can find more solutions to $p$ and consequently more nontrivial solutions to the original Yang-Mills equation on the 4d curved spacetimes with the metric \eqref{eq:4dMetricSphericalSymm-2}.

Let us discuss a few examples.
\begin{enumerate}[(1)]

\item Closed Friedmann universe:

It was firstly studied in \cite{Verbin-1}. The metric reads
\be
  \textrm{d}s^2 = R^2 (\chi) \left(\textrm{d}\chi^2 - a^2 \textrm{d}\Omega_3^2\right)\, ,
\ee
which is conformal to the Einstein static universe. The Euclidean version of this spacetime has the metric
\be
  \textrm{d}s^2 = \textrm{d}r^2 + a^2 \textrm{d}\Omega_3^2\, ,
\ee
which is conformal to
\be
  \textrm{d}s^2 = \frac{r^2}{a^2} \textrm{d}r^2 + r^2 \textrm{d}\Omega_3^2\, ,
\ee
and it is a special case of \eqref{eq:4dMetricSphericalSymm-1} with $f(r) = a$ or \eqref{eq:4dMetricSphericalSymm-2} with $h(r) = r^2 / a^2$. Using the Ansatz \eqref{eq:AnsatzCurved} we reproduce the instanton and the anti-instanton solutions found in \cite{Verbin-1} in the gauge $A_r = 0$:
\be
  p(r) = \frac{1}{1 + \textrm{exp} \left[\frac{2}{a} (r_0 \mp r) \right]}\, .
\ee

\item Euclidean AdS$_4$:

This case was discussed in \cite{Verbin-2}. One chooses the metric for the Euclidean AdS space to be
\be\label{eq:metricAdS4-1}
  \textrm{d}s^2 = a^2 \left(\textrm{d} \xi^2 + \textrm{sinh}^2 \xi\, \textrm{d}\Omega_3^2 \right)\, ,
\ee
which is a special case of \eqref{eq:4dMetricSphericalSymm-1} with $f(\xi) = a\, \textrm{sinh}\, \xi$. Defining $r \equiv a\, \textrm{sinh}\, \xi$, we can rewrite the metric \eqref{eq:metricAdS4-1} into the form of \eqref{eq:4dMetricSphericalSymm-2}:
\be\label{eq:metricAdS4-2}
  \textrm{d}s^2 = \frac{\textrm{d}r^2}{1 + \frac{r^2}{a^2}} + r^2\, \textrm{d}\Omega_3^2\, ,
\ee
i.e. $h(r) = (1 + r^2 / a^2)^{-1}$ in this case. Using the Ansatz \eqref{eq:AnsatzCurved} we reproduce the following nontrivial solution to Yang-Mills equation found in \cite{Verbin-2} with the gauge $A_r = 0$:
\be
  p(r) = \frac{r^2}{r^2 + \lambda^2 \left(a + \sqrt{r^2 + a^2} \right)^2}\, ,
\ee
where $\lambda$ is a free constant. This solution can be interpreted as the instanton solution on AdS$_4$, and it can be generalized to certain modified AdS spaces \cite{Nielsen}.

\item Lorentzian dS$_4$:

This case was considered recently in \cite{Lechtenfeld-1, Lechtenfeld-2, Lechtenfeld-3}. The metric for the Lorentzian dS$_4$ space is given by
\be\label{eq:dS4LorentzMetric-1}
  \textrm{d}s^2 = a^2 \left(- \textrm{d}\eta^2 + \textrm{cosh}^2 \eta\, \textrm{d}\Omega_3^2 \right)\, ,
\ee
which is different from the class \eqref{eq:4dMetricSphericalSymm-1}, because it has the Lorentzian signature. In order to apply the techniques discussed above to this case, we first perform a Wick rotation $\eta_E \equiv \textrm{i} \eta$ and obtain
\be\label{eq:dS4LorentzMetric-2}
  \textrm{d}s^2 = a^2 \left(\textrm{d}\eta_E^2 + \textrm{cos}^2 \eta_E\, \textrm{d}\Omega_3^2 \right)\, .
\ee
Defining $r \equiv a\, \textrm{cos}\, \eta_E$, we can rewrite the metric \eqref{eq:dS4LorentzMetric-2} into the form of \eqref{eq:4dMetricSphericalSymm-2}:
\be\label{eq:dS4LorentzMetric-3}
  \textrm{d}s^2 = \frac{\textrm{d}r^2}{1 - \frac{r^2}{a^2}} + r^2\, \textrm{d}\Omega_3^2\, ,
\ee
i.e. $h(r) = (1 - r^2 / a^2)^{-1}$ in this case. Using the Ansatz \eqref{eq:AnsatzCurved} we reproduce the following nontrivial solution to Yang-Mills equation:
\be\label{eq:SoldS4Lorentz}
  p(r) = \frac{r^2}{r^2 + \lambda^2 \left(a + \sqrt{a^2 - r^2} \right)^2}\, ,
\ee
where $r \in [0,\, a]$ in order for the solution to be real-valued.

There is an alternative way of finding solutions to Yang-Mills equation on the Lorentzian dS$_4$ \eqref{eq:dS4LorentzMetric-1}. Since the spatial part $S^3$ has an $SO(4) \cong SU(2)\times SU(2)$ isometry, we can identify one $SU(2)$ from the isometry group with the $SU(2)$ from the gauge group. Using the special property of $S^3$ (see Appendix~B of Ref.~\cite{SqS3}), we can construct an Ansatz obeying the form invariance condition \eqref{eq:FI} as follows:
\begin{align}
  A_\mu \textrm{d}x^\mu & = p(t)\, \left(U^{-1} \partial_\mu U \right)\, \textrm{d}x^\mu = p(t)\, \left(2 e_\mu^a T^a \right)\, \textrm{d}x^\mu \nonumber\\
  {} & = 2\, p(t)\, T^a\, e^a\, ,
\end{align}
where the generators $T^a$ are given by Eq.~\eqref{eq:generator}, and $e^a$ denote the vielbeins on $S^3$. Using this Ansatz, we reproduce the nontrivial solution to Yang-Mills equation found in Ref.~\cite{Lechtenfeld-1}:
\be
  p(t) = \frac{1}{2} \left[1 + \sqrt{2}\, \textrm{sech} \left(\sqrt{2} (t - t_0) \right) \right]\, ,
\ee
where $t_0$ is a constant.
\end{enumerate}


{\flushleft\bf 8.~Flat Space with Other Gauge Groups}

Besides the flat space and the curved space with the gauge group $SU(2)$, we can also consider the flat space with higher-rank gauge groups. For simplicity, let us focus on the $N$-dimensional Euclidean space with the gauge group $SO(N)$. For this case, since both the gauge group and the Lorentz group are $SO(N)$, one can write down the Ans\"atze similar to \eqref{eq:3DAnsatz} and \eqref{eq:4DAnsatz}. However, to simplify the discussion, an alternative Ansatz is more convenient to use in this case:
\be\label{eq:SO(N)Ansatz}
  A_\mu = q(r)\, M_{\mu\nu} x_\nu\, ,
\ee
where $r$ is formally defined in the same way as the flat 3d case \eqref{eq:Def_tau} but with the index $\mu$ running from $1$ to $N$, and $M_{\mu\nu}$ are the generators of $SO(N)$. One can prove that this Ansatz obeys the form invariance condition \eqref{eq:FI} (see Appendix~E.3 of Ref.~\cite{YM-1}). In some dimensions $N$, there can be more general Ans\"atze with other tensor structure satisfying the form invariance condition, e.g. $q(r)\, \epsilon_{\mu\nu\rho\sigma} M^{\nu\rho} x^\sigma$ \cite{Ma}, but the Ansatz \eqref{eq:SO(N)Ansatz} always provides a subsector of solutions to Yang-Mills equation. For this subsector, the solutions have the moduli from the size, from the Lorentz translations and from the gauge orientations, similar to the moduli corresponding to bosonic zero modes for the 4d $SU(2)$ instantons~\cite{tHooft:1976snw}. Hence, this sub-moduli space for the $N$-dimensional $SO(N)$ Yang-Mills theory has the dimension
\be\label{eq:ModuliDimension}
  1 + N + \frac{1}{2} N\, (N - 1) = \frac{N^2 + N + 2}{2}\, .
\ee

Inserting the Ansatz \eqref{eq:SO(N)Ansatz} into Yang-Mills equation, we obtain the simplified Yang-Mills equation as follows:
\be\label{eq:SimplifiedYM}
  r \frac{\textrm{d}}{\textrm{d}r} \left(\frac{1}{r} \frac{\textrm{d} q(r)}{\textrm{d}r} \right) + \frac{2 + N}{r}\, \frac{\textrm{d} q(r)}{\textrm{d}r} +  3 (2 - N)\, q^2 (r) + (2 - N) r^2\, q^3 (r) = 0\, .
\ee

One can consider the simplest 2d case. Since it corresponds to an Abelian gauge theory with the gauge group $U(1)$, Eq.~\eqref{eq:SimplifiedYM} has a simpler form
\be
  r \frac{\textrm{d}}{\textrm{d}r} \left(\frac{1}{r} \frac{\textrm{d} q(r)}{\textrm{d}r} \right) + \frac{4}{r}\, \frac{\textrm{d} q(r)}{\textrm{d}r} = 0\, ,
\ee
which has the solution
\be
  q(r) = \frac{c_1}{r^2} + c_2
\ee
with constants $c_1$ and $c_2$.

As one would expect, for all $N \in \mathbb{Z}_{\geq 3}$, there are always at least three solutions:
\begin{align}
  q (r)   = 0\, ,\,\, - \frac{1}{r^2}\, ,\,\, - \frac{2}{r^2}\, ,
\end{align}
which  for $N = 3$ correspond to the vacuum solution, the Wu-Yang monopole and the pure gauge solution respectively.

For $N=4$, we find the following exact solutions with nontrivial topology \cite{Ma}:
\begin{align}
  q (r) & = \frac{2}{2 - r^2}\, ,\\
  q (r) & = - \frac{4}{r^2 (2 + c r^2)}\quad \textrm{with a constant } c\, .
\end{align}
Following the general discussion \eqref{eq:ModuliDimension}, this class of solutions has the sub-moduli space of dimension $11$.

For $N=10$, we discover new exact solutions with nontrivial topology:
\begin{align}
  q (r) & = \frac{1}{1 - r^2}\, ,\\
  q (r) & = - \frac{2 + c r^2}{r^2 (1 + c r^2)}\quad \textrm{with a constant } c\, . 
\end{align}
Although these solutions to the 10d Yang-Mills equation are not the most general ones, to our best knowledge, they are new solutions and have not been discussed in the literature before. Following the general discussion \eqref{eq:ModuliDimension}, this class of solutions has the sub-moduli space of dimension $56$.

Besides the dimensions discussed above, we found many other numerical solutions yet their closed forms have not been established. To understand this fact and reveal more hidden mathematical structures of the solutions, further studies are needed.

{\flushleft\bf 9.~More General Cases}

From the discussions above, we demonstrate how our simple formalism could help us find exact solutions to Yang-Mills equation in various examples. It is not hard to generalize it to more general cases.

So far we have only discussed spherically symmetric solutions, for both flat spaces and curved isotropic spaces.  We consider these isotropic curved spaces, because they share the same symmetry with the corresponding flat spaces and can be treated in the same way, but the form invariance principle can be applied to less symmetric curved spaces. To go beyond the spherically symmetric case, we can glue multiple spherically symmetric solutions together to construct an approximate solution if the single-center solutions are put far away from each other. For the true exact multi-center solutions, we should generalize the exact multi-instanton solution \cite{multi-instanton} to curved spacetime, which breaks the spherical symmetry to cylindrical symmetry. This work will be presented in \cite{curved-multi-instanton}.

To apply the new formalism on more general spacetimes and gauge groups, we should first analyze the isometry group of the spacetime and compare it with the gauge group. Generally speaking, for a curved spacetime with an isometry group $H$ and for a gauge group $G$, we should find their subgroups $H' \subset H$ and $G' \subset G$, such that $H' \cong G'$. Since the rotation of the spacetime generated by $H'$ is isomorphic to the rotation of the gauge group space generated by $G'$, we can construct an Ansatz similar to the flat cases \eqref{eq:3DAnsatz} and \eqref{eq:4DAnsatz} with respect to the form invariance condition \eqref{eq:FI}. After that, we follow subsequent steps as described in general procedure. Due to the multiple choices of subgroups, interesting subclasses of solutions could be found.

For spacetimes without enough isometry, a similar formalism could be applied yet a more general form invariance condition~\eqref{eq:FIlocal} with local Lorentz transformations is needed.

\par
{\flushleft\bf 10.~Discussions}

A systematic way of finding exact solutions to classical Yang-Mills equation has been proposed. It is based on one of the Wightman axioms for quantum field theory, which is referred to as form invariance condition in this paper. Within this framework, all the previous known solutions in the literature are reproduced and new solutions are found.  As the first one in a series of papers, we only focus on the simplest cases with spherical symmetry in this paper, and the cases with less symmetries will be discussed in other papers that will appear soon. Moreover, we choose the curved spaces to be isotropic and have the same isometry group as the corresponding flat spaces to show that the spaces with the same symmetry allow the same treatments, because they obey the same form invariance condition, whether they are flat or curved is not crucial.

For the future research, we would like to classify all the solutions and study their moduli spaces. In particular, for the solutions with nontrivial topology that we found in this paper, we should be able to obtain complex saddle points discussed in Ref.~\cite{NekrasovNew}. We may also apply the approach discussed in this paper to YM equations with a source, e.g. a point particle carrying non-Abelian charge \cite{Kosyakov:1998qi, Kosyakov:2007qc} or a non-Abelian current \cite{Shirokov:2019eik}. The implications in (A)dS/CFT correspondence \cite{Maldacena, Strominger} from the nontrivial solutions on (A)dS spaces should also be explored in detail. According to the recent progress on the so-called double copy \cite{Tod, DoubleCopy-1, DoubleCopy-2, DoubleCopy-3, DoubleCopy-4}, we anticipate some nontrivial solutions in gravity corresponding to the solutions to Yang-Mills equation discussed in this paper. We can also generalize the formalism to the Yang-Mills theory coupled to scalars and try to reproduce the various nontrivial dilaton solutions discussed in the literature \cite{Actor:1979in}.

\par
{\flushleft\bf Acknowledgments}

We would like to thank Amjad Ashoorioon, Yu Jia, Olaf Lechtenfeld, David McGady, Miguel Montero, Vasily Pestun and Peng Zhao for many useful discussions. J.N.'s work was supported in part by the U.S. Department of Energy under grant DE-SC0007859 and by a Van Loo Postdoctoral Fellowship.

J.N. and Y.Q. contribute equally to this work.

\bibliographystyle{utphys}
\bibliography{YMcurved}

\end{document}